\documentclass[pra,twocolumn,showpacs,letterpaper,showpacs,superscriptaddress]{revtex4}
\usepackage{graphicx,amsmath,amssymb,amsfonts,latexsym,color,dcolumn,bm}

\begin{document}

\newcommand{\beq}{\begin{equation}}
\newcommand{\eeq}{  \end{equation}}
\newcommand{\bea}{\begin{eqnarray}}
\newcommand{\eea}{  \end{eqnarray}}
\newcommand{\bit}{\begin{itemize}}
\newcommand{\eit}{  \end{itemize}}

\title{Exact Casimir interaction between eccentric cylinders}

\author{D.A.R. Dalvit}
\affiliation{Theoretical Division, MS B213,Los Alamos National Laboratory,Los Alamos,NM 87545, USA}

\author{F.C. Lombardo}
\author{F.D. Mazzitelli}
\affiliation{Departamento de F\'{\i}sica J.J. Giambiagi,Facultad de Ciencias Exactas y 
Naturales,Universidad de Buenos Aires - Ciudad Universitaria, Pabell\'on I,1428 Buenos Aires, Argentina}

\author{R. Onofrio}
\affiliation{Department of Physics and Astronomy,Dartmouth College,6127 Wilder Laboratory,Hanover,NH 03755,USA}

\affiliation{Dipartimento di Fisica ``G. Galilei",Universit\`a di Padova,Via Marzolo 8,Padova 35131,Italy}

\date{\today}

\begin{abstract}

The Casimir force is the ultimate background in ongoing searches of extra-gravitational 
forces in the micrometer range. Eccentric cylinders offer favorable experimental conditions 
for such measurements as spurious gravitational and electrostatic effects can be minimized. 
Here we report on the evaluation of the exact Casimir interaction between perfectly conducting 
eccentric cylinders using a mode summation technique, and study different limiting cases of 
relevance for Casimir force measurements, with potential implications for the understanding 
of mechanical properties of nanotubes. 

\end{abstract}

\pacs{03.70.+k, 12.20.-m, 04.80.Cc}

\maketitle


As the size of physical systems is scaled down into the micrometer and submicrometer 
scales, macroscopic quantum effects become increasingly important. 
Forces attributable to the reshaping of quantum vacuum fluctuations under changes in geometrical 
boundary conditions, predicted almost sixty years ago by Casimir \cite{Casimir1948}, 
have been measured in recent years with increasing accuracy 
\cite{Lamoreaux1997,Mohideen1998,Ederth2000,Chan2001,Bressi2002,Decca2003}.
The fact that the magnitude and sign of the Casimir force depend both on the geometry and material 
structure of the boundaries paves the way to several opportunities and challenges 
for engineering mechanical structures above the nanoscale \cite{Chan2001prl}. 

The Casimir interaction for perfect metals has been exactly evaluated only for a limited
number of geometries, starting from the original parallel plate configuration \cite{Casimir1948}.
Until recently, for all non-planar geometries the Casimir force has been estimated using 
the so-called proximity-force approximation (PFA) \cite{Derjaguin1}, semiclassical 
\cite{Schaden00,Mazzitelli2003} and optical approximations \cite{Jaffe2004},
and numerical path-integral methods \cite{Gies2003}. This has originated  
a debate on the assessment of the accuracy of the measurements, a crucial
issue to establish reliable limits on extra-gravitational forces in the micrometer range.  
In the last months large violations of PFA for corrugated plates have
been reported \cite{Rodrigues2006}, and the exact Casimir interaction 
between a sphere in front of a plane and a cylinder in front of a plane has been computed 
\cite{Gies2006,Wirzba2006,Emig2006}.  
As first discussed in \cite{Dalvit2004}, the cylinder-plane configuration is intermediate 
between the plane-plane and sphere-plane geometries, offering easier parallelization than 
the former one, and a larger absolute signal than the latter one due to its extensivity in 
the length of the cylinder. A related experimental attempt aiming to measure temperature 
corrections to the Casimir force is under development \cite{BrownHayes2005}.

In this Rapid Communication we present the exact evaluation of the Casimir interaction for another 
geometry of experimental relevance consisting of two perfectly conducting eccentric cylinders. 
Although parallelism is as difficult as for the plane-plane geometry, this geometry offers several
experimental advantages. First, Gauss law dictates that the expected gravitational force is zero for 
any location of the inner cylinder, which allows for a null experiment when looking for 
intrinsically short-range extra-gravitational forces.
Second, the fact that the concentric configuration is an unstable equilibrium 
position \cite{Dalvit2004} 
opens the possibility of measuring the derivative of the force using closed-loop experiments. 
Finally, residual electrostatic charges on the surfaces can be exploited to maximize the parallelism 
between the cylinders looking at the minimum value of the resulting Coulomb force. 

Before embarking on the exact calculation of the Casimir energy for this geometry, let us recall
the result of the PFA. This is a simple, though uncontrolled way, of treating non-planar configurations, 
and is valid for surfaces whose separation is much smaller than typical local curvatures. 
For two very long eccentric cylinders of radii $a<b$, length $L$, and eccentricity $\epsilon$
(see Fig. 1), the PFA approximation for the non-concentric Casimir energy reads 
$E_{\rm PFA} = - \pi^3 \hbar c L \epsilon^2 / 120 a^4 (\alpha-1)^5$, valid when 
$\alpha \equiv b/a \rightarrow 1$ and for small eccentricity $\delta \equiv \epsilon/a \ll 1$. 
In the limit of large eccentricity ($\epsilon \approx b-a$) PFA predicts a behavior similar to that of 
a cylinder in front of a plane \cite{Dalvit2004}. 
The geometrical dimensionless parameters $\alpha$ and $\delta$ fully characterize the eccentric 
cylinders configuration.

\begin{figure}[t]
\setlength{\unitlength}{1cm}
\begin{center}
\scalebox{0.95}[0.95]{%
\includegraphics*[width=8cm]{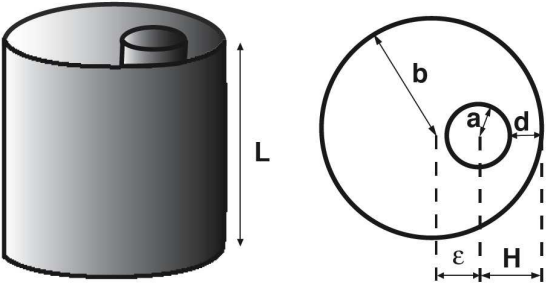}}
\end{center}
\caption{Geometrical configuration studied in this paper. Two perfectly conducting eccentric
cylinders of radii $a<b$, length $L$, and eccentricity $\epsilon$ interact via the Casimir force.
The equilibrium position at $\epsilon=0$ is unstable - any small perturbation will make
the cylinders snap into each other.}
\label{fig1}
\end{figure}

In order to go beyond the PFA result, we start by expressing the Casimir energy as  
$E=(\hbar/2) \sum_p (\omega_p - \tilde{\omega}_p)$, where $\omega_p$ are the eigenfrequencies
of the electromagnetic field satisfying perfect conductor boundary conditions on the cylindrical
surfaces, and $\tilde{\omega}_p$ are the corresponding ones to the reference vacuum (cylinders
at infinite separation). In cylindrical coordinates, the eigenmodes are $h_{n,k_z} = R_n(r,\theta)
\exp[ -i( \omega_{n,k_z} t - k_z z)]$, where $\omega_p = \omega_{n,k_z} = \sqrt{k_z^2 + \lambda_n^2}$,
and $R_n$ ($\lambda_n$) are the eigenfunctions (eigenvalues) of the 2D Helmholtz equation.
Using the argument theorem the sum over eigenmodes can be written as an integral over the complex
plane, with an exponential cutoff for regularization. In order to determine the part of the energy 
that depends on the separation between the two cylinders it is convenient to subtract the self-energies
of the two isolated cylinders, $E_{12} (a,b,\epsilon )= E - E_1(a) - E_1(b)$. Then the divergencies in $E$
are cancelled out by those ones in $E_1(a)$ and $E_1(b)$, and the final result for the interaction
energy is
\begin{equation}
E_{12}(a,b,\epsilon ) = \frac{\hbar c L}{4 \pi} \int_0^{\infty} dy \; y \; \log M(i y).
\label{exactenergy}
\end{equation}
Here $M = (F/F_{\infty}) / [(F_1(\infty) / F_1(a)) \; (F_1(\infty)/F_1(b)) ]$. 
The function $F$ is analytic and it vanishes at all the eigenvalues $\lambda_n$ ($F_{\infty}$, 
at  $\tilde{\lambda}_n$), and, similarly, $F_1$ vanishes for all eigenvalues for the isolated 
cylinders. The function $M$ is the ratio between a function corresponding to the actual geometrical 
configuration and one with the conducting cylinders far away from each other. 
As this last configuration is not univocally defined, we use this freedom to choose a particular one
that simplifies the calculation. It is convenient to subtract a configuration of two cylinders
with very large and very different radii, while keeping the same eccentricity of the original configuration.
Eq. (\ref{exactenergy}) is valid for two perfect conductors of any shape, as long as there is 
translational invariance along the $z$ axis. 

The solution of the Helmholtz equation in the annulus region between eccentric cylinders has been 
considered in the framework of classical electrodynamics and fluid dynamics \cite{Singh1984,Balseiro1950}. 
The eigenfrequencies for Dirichlet  boundary conditions (TM modes) and for Neumann boundary conditions 
(TE modes) are given by the zeros of the determinants of the non-diagonal matrices
\begin{eqnarray}
Q^{\rm TM}_{mn}&=& 
\left[J_n(\lambda a) N_m(\lambda b) - J_m(\lambda b) N_n(\lambda a) \right]
J_{n-m}(\lambda \epsilon) ,  \nonumber \\
Q^{\rm TE}_{mn}&=& 
\left[J'_n(\lambda a)N'_m(\lambda b) - J'_m(\lambda b) N'_n(\lambda a)\right]
J_{n-m}(\lambda \epsilon) , \nonumber 
\end{eqnarray}
where $J_n$ and $N_n$ are Bessel functions of the first kind.  
The function $M$ can be written as $M = M^{\rm TE} M^{\rm TM}$, where 
$M^{\rm TM}$ is built with ($R$ being a very large radius)
\begin{eqnarray}
&&F^{\rm TM} = {\rm det} 
\left[ Q^{\rm TM}(a,b,\epsilon)Q^{\rm TM}(b,R,0) \right] \prod_n J_n(\lambda a), \nonumber \\
&&F_1^{\rm TM}(a) = {\rm det} \left[ Q^{\rm TM}(a,R,0) \right] \prod_n J_n(\lambda a), 
\end{eqnarray}
Similar expressions hold for $M^{\rm TE}$.

The Casimir energy can be decomposed as a sum of TE and TM contributions
\begin{equation}
E_{12} = \frac{\hbar c L}{4 \pi a^2} \int_0^{\infty} d\beta \beta 
\left[ \log M^{\rm TE} \left( \frac{i\beta}{a} \right) 
+ \log M^{\rm TM} \left( \frac{i\beta}{a} \right) \right]
\label{exact}
\end{equation}
with $M^{\rm TE,TM}(\frac{i\beta}{a}) = {\rm det} [ \delta_{np} - A^{\rm TE,TM}_{np}]$. 
The  non-diagonal matrices $A^{\rm TE}_{np}$ and $A^{\rm TM}_{np}$ are 
\begin{eqnarray}
A_{np}^{\rm TM}  &=& \frac{I_n(\beta)}{K_n(\beta)} \sum_m \frac{K_m(\alpha\beta)}{I_m(\alpha\beta)} I_{m-n}(\beta \delta) I_{m-p}(\beta \delta) , \nonumber \\
A_{np}^{\rm TE}  &=& \frac{I'_n(\beta)}{K'_n(\beta)} \sum_m \frac{K'_m(\alpha \beta)}{I'_m(\alpha \beta)} I_{m-n}(\beta \delta) I_{m-p}(\beta \delta) \nonumber .
\end{eqnarray}
Here $I_n$ and $K_n$ are modified Bessel functions of the first kind.
The determinants are taken with respect to the integer indices 
$n,p = -\infty, \ldots, \infty$, and the integer index $m$ runs from $-\infty$ to $\infty$. 
Eq.(\ref{exact}) is the exact formula for the interaction Casimir energy between eccentric cylinders. 
This formula coincides with the known result for the Casimir energy for 
concentric cylinders ($\epsilon = 0$). As $I_{n-m}(0) = \delta_{nm}$, in this particular case 
the matrices $A_{np}^{{\rm TE,TM}}$ become diagonal \cite{Mazzitelli2003,Saharian2006}. 

\begin{figure}[t]
\setlength{\unitlength}{1cm}
\begin{center}
\scalebox{0.95}[0.95]{%
\includegraphics*[width=8cm]{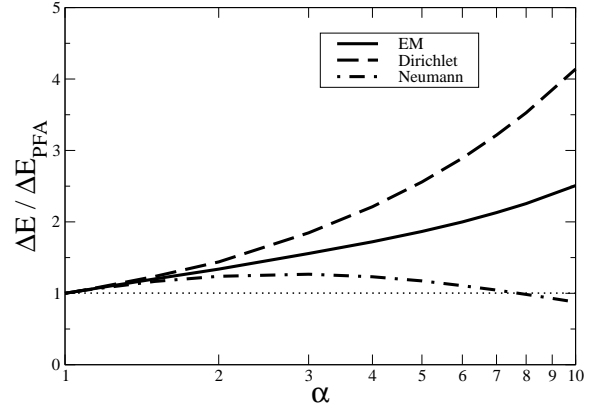}}
\end{center}
\caption{Ratio of the exact and PFA Casimir interaction energy differences 
$\Delta E = E_{12} - E^{\rm cc}_{12}$ between eccentric ($E_{12}$) and concentric 
($E^{\rm cc}_{12}$) cylinders in the limit of small eccentricity $\epsilon \ll a$. The curve EM
denotes the full electromagnetic Casimir energy.}
\label{fig2}
\end{figure}

We can also obtain the exact Casimir interaction energy for the cylinder-plane 
configuration \cite{Emig2006}
as a limiting case of our exact results for eccentric cylinders Eq.(\ref{exact}). Indeed, 
the eccentric cylinder configuration tends to the cylinder-plane configuration for
large  values of both the eccentricity $\epsilon$ and the radius $b$ of the outer cylinder, keeping the 
radius $a$ of the inner cylinder and the distance $d$ between the cylinders fixed. 
Using the addition theorem  and uniform expansions for Bessel functions 
it can be proved that, for $x \gg h$, 
\begin{eqnarray}
\sum_m \frac{K_m(x+h)}{I_m(x+h)} I_{n-m}(x) I_{p-m}(x) \approx K_{n+p}(2h) , \nonumber \\
\sum_m \frac{K'_m(x+h)}{I'_m(x+h)} I_{n-m}(x) I_{p-m}(x) \approx - K_{n+p}(2h) . \nonumber
\end{eqnarray}
Using  these equations (with $x\equiv \beta \epsilon/a$ and $h\equiv \beta H/a$)
in our exact formula the known result for the Casimir energy in the cylinder-plane configuration 
is obtained \cite{Emig2006}
\begin{eqnarray}
A_{np}^{\rm TM, c-p} &=& \frac{I_n(\beta)}{K_n(\beta)} \; K_{n+p}(2 \beta H/a) , \nonumber \\
A_{np}^{\rm TE, c-p} &=& - \frac{I'_n(\beta)}{K'_n(\beta)} \; K_{n+p}(2 \beta H/a) . \nonumber
\end{eqnarray}

\begin{figure}[t]
\setlength{\unitlength}{1cm}
\begin{center}
\scalebox{0.95}[0.95]{%
\includegraphics*[width=8cm]{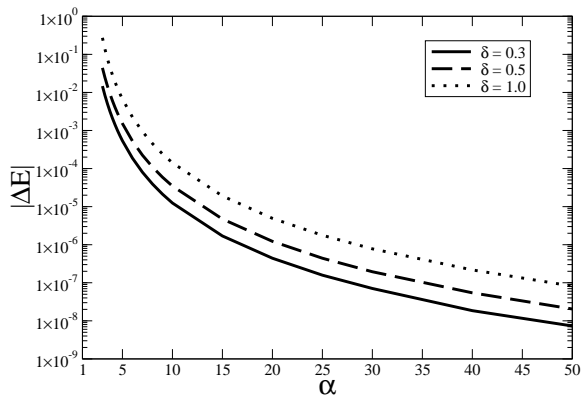}}
\end{center}
\caption{Exact interaction energy difference $|\Delta E|$ as a function of $\alpha$ for 
different values of $\delta$. Energies are measured in units of $\hbar c L / 4 \pi a^2$. These 
results interpolate between the $(\alpha - 1)^{-5}$ behavior for $\alpha\rightarrow 1$, and 
the $(\alpha^4 \log\alpha)^{-1}$ behavior for $\alpha \gg 1$.}
\label{fig3}
\end{figure}

We now focus on quasi-concentric cylinders, when the eccentricity is small as compared to the 
radius of the inner cylinder, i.e. $\delta \ll 1$. The ratio between the radii, $\alpha=b/a$, 
need not be close to unity, so that PFA is in general not valid in this configuration - as 
only for $\alpha \rightarrow 1$ do we expect to recover PFA. The zeroth order $\epsilon=0$ corresponds
to the concentric case, where the matrices $A_{np}^{\rm TE,TM}$ are diagonal. This case was studied
in \cite{Mazzitelli2003} by means of exact and semiclassical treatments, and it was shown that the PFA 
concentric energy goes as $E^{\rm cc}_{12} = - \hbar c \pi^3 L / 360 a^2 (\alpha-1)^3$.  
Obviously, symmetry arguments imply that there is no net Casimir force between the cylinders 
in the concentric configuration. To next order the correction to the energy, 
$\Delta E = E_{12} - E^{\rm cc}_{12}$, depends on $\epsilon^2$, leading to an unstable 
equilibrium position for the concentric geometry. The behavior of the Bessel functions 
for small eccentricity, $I_{m-n}(z) \simeq (z)^{n-m}$, suggests that one should use the 
tridiagonal version of the matrices $A_{np}^{\rm TE,TM}$ considering only elements with 
$p=n$ and $p = n \pm 1$. Expanding the determinants to order $O(\epsilon^2)$, one can write 
the TM contribution as
\begin{eqnarray}
\Delta E^{\rm TM} &=& - \frac{\hbar c L \epsilon^2}{4 \pi a^4} \sum_{n=-\infty}^{\infty} \int_0^{\infty} d\beta \; 
\beta^3 \;
\frac{1}{1-{\cal D}^{\rm cc}_n } \nonumber \\
& & \times \left[ {\cal D}_n  + \frac{ {\cal N}_n }{1- {\cal D}^{\rm cc}_{n+1}}
\right], 
\label{exactsmalleccentricity}
\end{eqnarray}
where ${\cal D}^{\rm cc}_n = I_n(\beta) K_n(\alpha \beta) / K_n(\beta) I_n(\alpha \beta)$ 
is the $\epsilon=0$, diagonal TM contribution, and ${\cal D}_n$ and ${\cal N}_n$ are 
the $O(\epsilon^2)$ diagonal and non-diagonal TM contributions, respectively. 
They read
\begin{eqnarray}
{\cal D}_n &=& \frac{ {\cal D}^{\rm cc}_n}{2} + \frac{I_n(\beta)}{4 K_n(\beta)} \left[
\frac{K_{n-1}(\alpha\beta)}{I_{n-1}(\alpha \beta)} + 
\frac{K_{n+1}(\alpha \beta)}{I_{n+1}(\alpha \beta)} \right] , \nonumber \\
{\cal N}_n &=& \frac{I_n(\beta) I_{n+1}(\beta)}{4 K_n(\beta) K_{n+1}(\beta)} \left[
\frac{K_{n}(\alpha \beta)}{I_{n}(\alpha \beta)} + \frac{K_{n+1}(\alpha \beta)}{I_{n+1}(\alpha \beta)} \right]^2 .
\nonumber
\end{eqnarray}
A similar expression holds for the TE contribution, with the Bessel functions replaced by their
derivatives. Eq.(\ref{exactsmalleccentricity}) and the corresponding TE one are exact expressions 
for the Casimir energy difference between eccentric and concentric cases in the limit  $\epsilon \ll a$. 

\begin{figure}[t]
\setlength{\unitlength}{1cm}
\begin{center}
\scalebox{0.95}[0.95]{%
\includegraphics*[width=8cm]{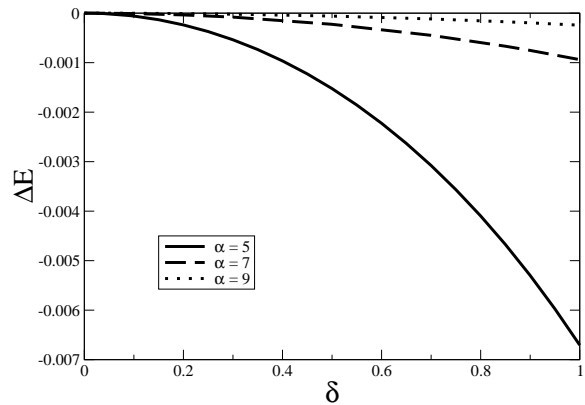}}
\end{center}
\caption{Exact interaction energy difference $\Delta E$ as a function of $\delta$ for 
different values of $\alpha$. Energies are measured in units of $\hbar c L / 4 \pi a^2$. The maximum 
at $\delta = 0$ shows the instability of the equilibrium point.}
\label{fig14}
\end{figure}

The PFA limit $\Delta E^{\rm TE}_{\rm PFA} = \Delta E^{\rm TM}_{\rm PFA} = 
\Delta E^{\rm EM}_{\rm PFA} /2 = - \pi^3 \hbar c L \epsilon^2 / 240 a^4 (\alpha-1)^5$
can be obtained from Eq.(\ref{exactsmalleccentricity}) considering $\alpha \rightarrow 1$.
In this limit the leading contribution arises from large values of the summation index $n$, 
for which the use of asymptotic uniform expansions of the Bessel functions is in order. 
Fig. 2 depicts the ratio of the exact Casimir energy $\Delta E$ and the PFA limit  for the almost
concentric cylinders configuration. As evident from the figure, PFA agrees with the exact result 
at a few percent level only for $\alpha$ very close to unity, and then it noticeably departs from 
the PFA prediction. In the opposite limit, $\alpha \rightarrow \infty$, the Dirichlet contribution 
is much larger than the Neumann one, and the integral and sum in Eq.(\ref{exactsmalleccentricity}) is 
dominated by the $n = 0$ term. The asymptotic result for the energy is
\begin{eqnarray}
\Delta E_\infty &=& - \frac{\hbar c L \epsilon^2}{8 \pi a^4 \alpha^4 \log \alpha} 
\int_0^{\infty} d\beta \beta^3 \left[ \frac{K_0(\beta)}{I_0(\beta)} + \frac{K_1(\beta)}{I_1(\beta)} \right] \nonumber \\
& \approx & - \frac{3.3348 \hbar c L \epsilon^2}{8 \pi a^4 \alpha^4 \log \alpha} .\label{largealpha}
\end{eqnarray}
This equation is valid when $\log\alpha \gg 1$. From Eq.(\ref{largealpha}) we see that 
the force between cylinders in the limit $a$, $\epsilon \ll b$ is proportional to 
$L\epsilon/b^4 \ln (b/a)$. The weak logarithmic dependence on $\alpha$ is 
characteristic of the cylindrical geometry. A similar weak decay at large 
distances has been found for the cylinder-plane geometry \cite{Emig2006}.

Next we consider arbitrary values of the eccentricity. In this case we need to perform a
numerical evaluation of the determinants in Eq.(\ref{exact}). We find that as $\alpha$ 
approaches smaller values, larger matrices are needed for ensuring convergence. 
Moreover, for increasing values of $\delta$ it is necessary to include more terms in 
the series defining the coefficients $A_{np}^{\rm TE,TM}$. In Fig. 3 we plot the interaction energy 
difference $|\Delta E|$ as a function of $\alpha$ for different values of $\delta$. These 
numerical results interpolate between the PFA and the asymptotic behavior for large $\alpha$, 
beyond the quasi-concentric limit. Fig. 4 shows the complementary information, with the 
Casimir energy as a function of $\delta$ for various values of $\alpha$, showing 
explicitly the instability of the concentric equilibrium position. 

The exact evaluation of the Casimir force between parallel eccentric cylinders obtained here lends 
itself to a variety of applications for experiments implementing this geometry in the micrometer 
and the nanometer scales. Firstly, the accurate knowledge of the Casimir force in this 
configuration allows to look for extra-gravitational forces while the usual gravitational 
force, apart from border effects, is cancelled. The cancellation of the Newtonian gravitational 
force is also common to the parallel plate geometry if a dynamical measurement technique is adopted for the 
latter configuration, however in addition the concentric cylinder case has better shielding from the 
electrostatic force due to spurious charges, allowing for a null experiment with respect to 
both Newtonian and Coulombian background forces even with a static measurement technique.  
This suggests a micrometer version of experiments performed using the concentric 
cylinders configuration and a torsional balance to test the inverse-square 
gravitational law in the cm range \cite{Spero}.
In particular, in the case of a repulsive Yukawian force, one can envisage a situation 
where the unstable equilibrium due to the Casimir force is balanced or overcome by the 
former, and even the qualitative observation of mechanical stability in the concentric 
configuration will establish the existence of a new force. 
Secondly, the measurement of the Casimir force in various configurations involving 
cylinders (with the first example provided in \cite{Ederth2000} for crossed cylinders) 
is interesting in itself as one can make reliable tests of the PFA approximation versus 
both the exact solution and the actual experimental outcome. We expect that 
the case of two parallel cylinders with distance larger than the sum of their radii 
will experience stronger deviations between PFA and the exact solution. 
This case could be analyzed using the approach presented here. 
Thirdly, our results could lead to a quantitative explanation, once 
finite conductivity and temperature effects will be taken into account, 
for the observed easiness to bend laterally multiwall nanotubes. 
The Casimir force and its 
non-retarded counterpart \cite{Hertel} provide a natural mechanism to allow for 
instability from the perfectly concentric situation of nanotubes, and 
could also play a role in the observed decrease of the effective bending modulus 
of nanotubes with large (tens to hundreds nm) radii \cite{Dresselhaus}, 
and their fragmentation \cite{Wood}.

The work of F.C.L. and F.D.M. is supported by UBA, Conicet and ANPCyT (Argentina). 

\end{document}